\documentclass[sprocl]{article}

\usepackage{epsfig}
\usepackage{sprocl}

 \def\be{\begin{equation}}
\def\ee{\end{equation}} \def\bea{\begin{eqnarray}}
\def\eea{\end{eqnarray}}

\def\shop{[\raisebox{.3ex}{$\mathcal{S}$} \hskip-0.5em
            \raisebox{-.3ex}{$\mathcal{L}$}]} 

\begin{document}

\title{PRECISION STUDIES AT A FUTURE LINEAR COLLIDER}

\author{DAVID J. MILLER}

\address{Physics and Astronomy, University College London,\\ 
         Gower Street, London WC1E 6BT, UK \\ 
         email: djm@hep.ucl.ac.uk}

\maketitle\abstracts{A future linear collider will complement the
programme of the Large Hadron Collider, especially for precision
physics.  The RADCOR community has an important part to play in
refining the predictions for rates and processes in Electroweak,
Higgs, SUSY, top and QCD physics, for $e^+ e^-$, $e^- e^-$, $\gamma
\gamma$ and $\gamma e$ collisions.}

\vspace{0.1cm}
\begin{center}
  Invited talk at RADCOR, Barcelona, Espa\~na,  8--12 Sept 1998
\end{center}
\vspace{0.2cm}

\section{Introduction}
Linear colliders in the range from 300 GeV to some Tev will tackle
many more topics in precision physics than can be dealt with in this
short talk.  They will also set out to discover new effects and new
particles.  Studies of the whole physics programme are under way in
Europe~\cite{ECFAD}, in Asia~\cite{asia} and America~\cite{america},
with Worldwide co-ordination~\cite{wcoord}.  The participants will get
together to present results at the Linear Collider Workshop (LCWS) at
Sitges, just down the road from here, next April.  Those who want to
know more should come back to Barcelona then.  But enough is already
known about the goals of such machines from previous
studies~\cite{morioka,waikoloa,saariselka,desy123} to motivate a
highlighted shopping list of topics where input from the RADCOR
community will be particularly important.  Some of them are picked out
in the text of this review with the shopping-list symbol ``\shop''.  A
lot of phenomenological Ph.D. projects will need to be completed
before we are ready to exploit the full physics potential of a linear
collider.

\begin{figure}
 \begin{center}
   \epsfig{file=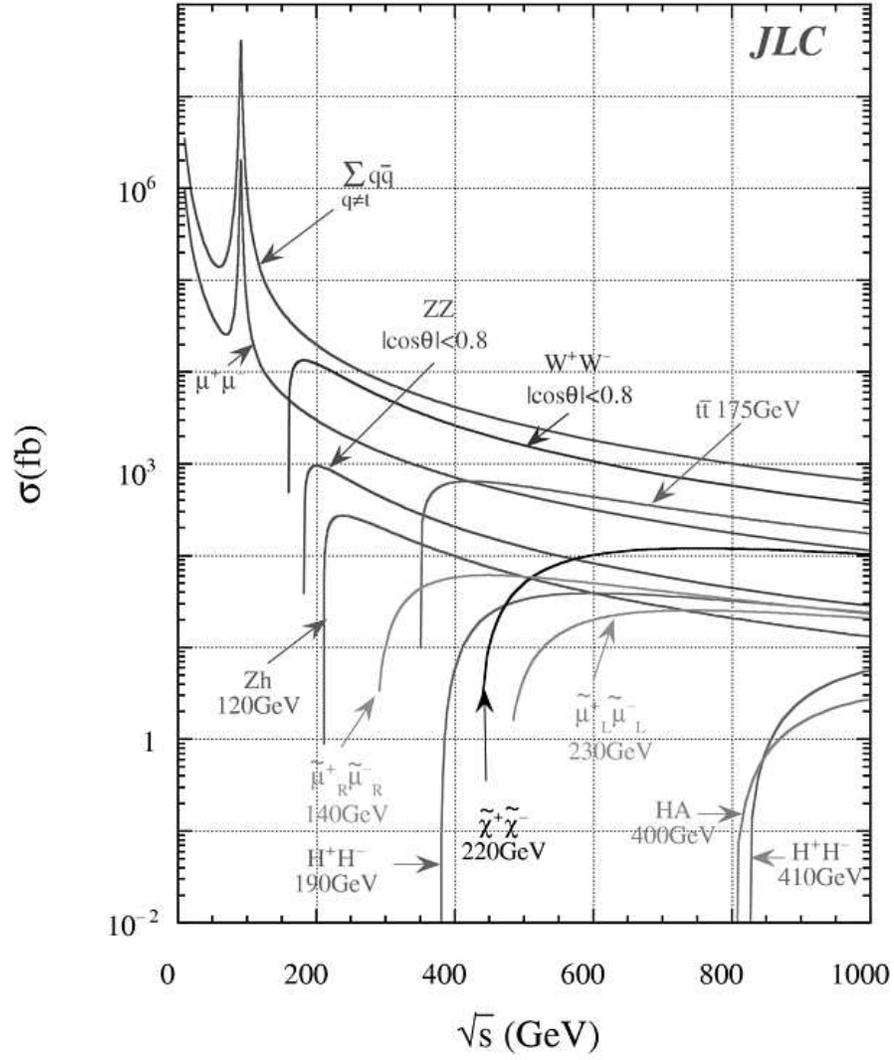,width=1.0\linewidth} 
 
  \vspace{-0.1cm}
  \caption{\label{fig:rates} 
        Cross sections for some interesting channels.}
 \end{center}
\end{figure}

There are likely to be two stages of machine-building over two
decades; perhaps following one-another in the same tunnel, or perhaps
on different continents.  The first will cover the range from 300 GeV,
near the top threshold, to about 1 TeV. Such a machine might also be
able to revisit the $Z^0$ region and the $W^+ W^-$ threshold with high
luminosity and polarised beams, if the physics interest is strong
enough to justify the extra engineering required.  The luminosity will
be in the range~\cite{techrev,tesla} from $5\cdot 10^{33}$ to
$5\cdot 10^{34}$, giving useful event samples for a year 
of running between $50fb^{-1}$ to $500fb^{-1}$ in a given small
energy range.  Figure~\ref{fig:rates} shows cross sections for some of
the interesting channels.  Note that many of them are comparable with
the $e^+ e^- \rightarrow \mu^+ \mu^-$ rate; ``a few units of R'' as we
say, where R is the ratio of the cross-section for a given process to
the QED dimuon cross section.  The option of running with $e^- e^-$ as
well as $e^+ e^-$ will be relatively easy to provide.  There are also
possibilities for special Compton-backscattered photon beams at up to
$80\%$ of full beam energy~\cite{gamgam,djmorioka} and almost the full
luminosity, giving a much better source of $e \gamma $ and $\gamma
\gamma $ events than ever before. These will require the solution of
some tricky, but probably not insuperable, technical problems.  The
RADCOR community should be involved in making the physics case to
justify these developments \shop.

The second-stage machine, in the energy-range from 1.5 to 4 TeV, will
be particularly important if it turns out that there is no narrow 
Higgs boson below 600 GeV, and/or if new physics (such as SUSY) is 
discovered at the LHC or at the first-stage collider, with a predicted 
spectrum of particles which are too massive to be seen by those 
machines.  The luminosity scaling law $\mathcal{L} \propto s$ will 
compensate the fall in cross-sections with rising energy, so long as
the ever-smaller beamspots can be kept in collision.  On the time
scale for building the second-stage linear collider it is possible
that a muon collider could be a feasible and cost effective
alternative~\cite{fermiweb,Keilrol,Blondel}.
 
There are four active design programmes for a linear collider, based
on four major laboratories.  SLAC and KEK are collaborating on an
X-band normally conducting machine~\cite{Xbandweb,JLC}.  A serious
technical proposal for this may be submitted within three years.  DESY
(and collaborators) have the rival superconducting TESLA
design~\cite{tesla} which has recently been upgraded to predict a
luminosity approaching $5\cdot 10^{34}$ at 500 GeV C. of M.  They also plan
an early technical proposal.  KEK also has a normally conducting
C-band design under study.  And at CERN successful first steps have
been made to prove the feasibility of a two-beam accelerator called
CLIC~\cite{CLIC} which looks like the best prospect for the
second-stage linear collider.

\section{The Detectors}
The demands of physics at the first-stage machine will be a little
more stringent than at LEP2, but nowhere near as difficult as at the
LHC.  Some of the most interesting channels (e.g. $t \bar{t}
\rightarrow W^+ b W^- \bar{b}$) will have six jets instead of two or
four, and better resolution of energy flow in the calorimeters will be
needed to separate them.  The beam pipe can have a much smaller radius
than at LEP, 1 or 2 cm compared with 10 cm, which will allow much
better resolution of beauty decays.  If the radius is brought down to
1 cm then the efficiency for identifying charm will be greatly
enhanced at the cost of a longer and more expensive final-focus
insertion in the linac.  Phenomenologists must be involved in
developing the case for this \shop.

After all the R\&D on radiation-hard, fast, high density detector 
technology for the Large Hadron Collider there are many new techniques 
available for the Linear Collider detector.  A number of contrasting 
strategies are being investigated.  One radical approach, discussed in 
the Snowmass study in 1996~\cite{snowmass}, is to build a compact
silicon tracker in a 4 tesla solenoid field and start the
electromagnetic calorimetry within a meter from the intersection
point.  European~\cite{tesla} and Japanese~\cite{jlc} studies have
gone for much larger tracking volumes with gaseous detectors (TPC or
Jet Chamber).  There is a debate at the moment on the need for special
cerenkov or transition-radiation devices for particle identification.
The developing consensus is to omit them.  The choice between the
different detector options will be made by looking at their
performance on a set of ``reference reactions'' which will include
both the most important channels and those which put the heaviest
demands on detector performance.  The RADCOR community is already
involved, helping choose the channels and writing Monte Carlo
generators; but experimenters are never satisfied with the available
generators so much more work will be needed \shop.

\begin{figure}[htb]
 \begin{center}
   \epsfig{file=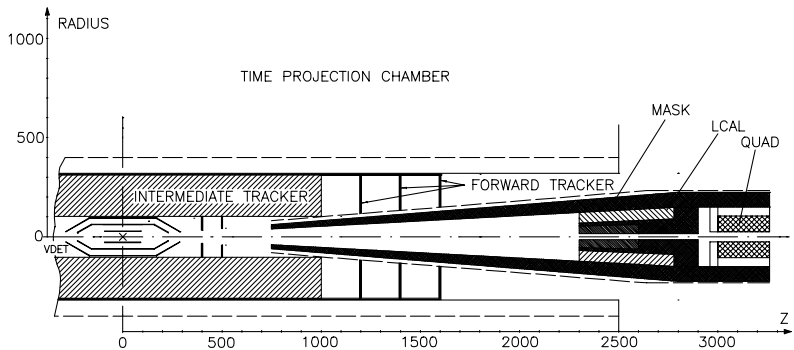,width=1.0\linewidth} 
 
  \vspace{-0.7cm}
  \caption{\label{fig:mask} 
     Schematic layout of the inner region of a linear collider  detector.}
 \end{center}
\end{figure}

Because of the electromagnetic background at a future linear collider 
the parts of the detector close to the forward beam directions will be 
much less effective than at LEP.  This will make the detector less 
hermetic than the best LEP detectors and will undermine some new 
particle searches.  The reasons are inescapable.  The only way of 
achieving high luminosity with beams making a single pass is to pinch
them down to a few nm in the vertical direction, but this gives rise
to new electromagnetic effects including beamstrahlung, beam 
disruption and copious pair production by the high energy leptons of 
one beam interacting with the electromagnetic field of the whole 
opposing bunch.  The disrupted beams can be accommodated by careful 
beamline design, but the soft pairs come out at angles of tens of 
degrees and contain a total of many TeV of energy per beam crossing. 
The electrons and positrons are trapped by the solenoid field and go 
in the forward direction.  When they collide with small-angle forward 
detectors or with the final-focus quadrupole magnets they produce 
large numbers of backscattered soft gamma rays and x-rays.  To stop 
this background from swamping the main tracking detectors all detector 
designs incorporate a thick conical tungsten mask which cuts off the 
region within approximately 100 mr from the beam direction and which 
projects some 10s of cm from the intersection point, see
Figure~\ref{fig:mask}.  Outside the mask detectors should operate in a
relatively quiet environment.  Inside the mask any detectors will need
to be radiation-hard, and will have to identify wanted signals on top
of the wash of soft electromagnetic background energy.  Some of us
are looking at ways of incorporating a minimal quantity of sampling
material in the mask itself so that the solid angle which it covers is
not totally dead.

\section{Top Quark Physics}
\label{sec-top}
The scan across the threshold for $t \bar{t}$ production is one of the
first jobs to be done, and one requiring the highest precision.  The
first goal is to measure the mass of the top quark.  It may also be
possible to deduce something indirectly about the width.  The
experimental systematic error could be as low as $\pm 120$
MeV~\cite{tesla}, with a ten-point scan of $50fb^{-1}$ integrated
luminosity; if the effect of beamstrahlung upon the luminosity
spectrum can be adequately monitored, see section~\ref{sec:luspec}
below \shop.  The short lifetime of the top quark acts as a cutoff
which permits a perturbative QCD calculation of the excitation curve.
But recent work~\cite{melnikov,hoang} has shown that the NLO and NNLO
corrections move the position of the step on the curve sideways by
about 1 GeV and change the predicted height of the step in cross
section by tens of percent.
The eventual
precision on the top mass is now a problem for the theorists, so this
item is firmly on the shopping list \shop.

The other great goal in $t \bar{t}$ studies is to measure the Yukawa
coupling $\lambda$ of the higgs boson to the top current.  In the
early 1990s, before the top quark mass was known, we wondered if it
might be possible to measure $\lambda$ from the height of the step in
the threshold excitation curve; for a very heavy top quark the higgs
loop would become significant compared with the leading order gluon
loop.  But now we know the top mass is too light for any such effect
to be detected (and, for the moment, we cannot calculate the QCD
contribution to sufficient precision).

\begin{figure}[htb]
 \vspace{-0.5cm}
 \begin{center}
   \epsfig{file=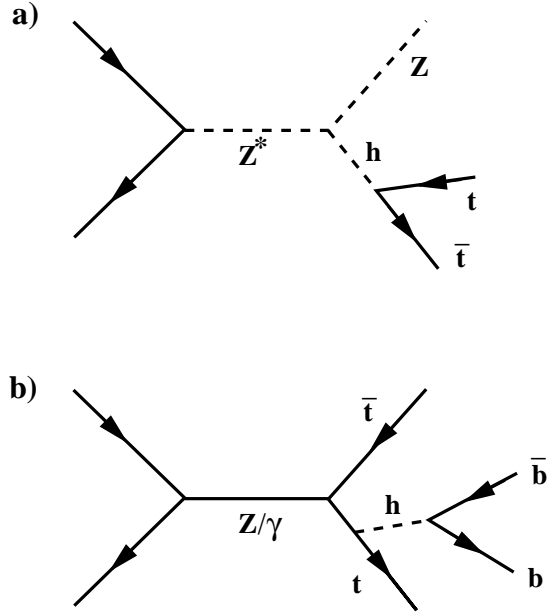,width=0.6\linewidth} 
 
  \vspace{0.2cm}
  \caption{\label{fig:Yukawa}
     Ways of measuring the Higgs boson Yukawa coupling for a) heavy
     Higgs, b) light Higgs.}
 \end{center}
\end{figure}

But the Yukawa coupling should be measurable by more direct methods,
especially if the high luminosity promised by TESLA can be delivered.
If the higgs mass is more than twice the top mass then we search for
$h\rightarrow t \bar{t}$, Figure~\ref{fig:Yukawa}a.  Old studies by
Fujii~\cite{Fujii} (with $m_h =300$ GeV and $m_t =130$ GeV!)
suggested that a clear signal would be seen with $60 fb^{-1}$ at
$\sqrt{s} =600$ GeV.  These need updating \shop\ to take account of
the known top mass and the possibility of higher luminosity.  If the 
higgs mass is less than the top mass there will be a significant rate 
for the radiation of higgs bosons from an outgoing top quarks in $t 
\bar{t}$ production, see Figure~\ref{fig:Yukawa}b.  Juste's old
study~\cite{juste} predicted a signal of 48 such events, after cuts,
with background of 29 events, in $50 fb^-1$ with $\sqrt{s} = 750$
GeV.  At that level he suggested that a significant measurement of
$\lambda$ would be ``barely possible''.  But with 10 times the
luminosity the situation would be very different.  And a new
calculation~\cite{dittm} of corrections to Figure~\ref{fig:Yukawa}b
shows a significant enhancement to the rate for lower values of
$\sqrt{s} \simeq 500$ GeV; another process already on the shopping
list \shop.
 
\section{Monitoring the Luminosity Spectrum}

\label{sec:luspec}
Measuring luminosity at a Future Linear Collider will be a much more
difficult task than at LEP or SLC, but precision physics at the $t
\bar{t}$ threshold, or at any SUSY particle thresholds, will depend on
it.  If the mass of the top quark is to be measured to $\pm175$ MeV
(say) then the resolution on the collision energy has to be better
than 1/1000.  The big problem is beamstrahlung, the loss of energy by
electrons in one beam from the intense electromagnetic field of the
opposite bunch.  This causes a variable amount of radiative energy
loss before the main collision takes place.  There can also be a
significant spread of energy in the linac beams, much more than in a
circular collider, though this can be reduced in most linac designs --
sometimes at the expense of reduced luminosity.  For any period of
running we will need to determine the luminosity spectrum of the
collisions, not just the integrated luminosity.  And no experimenter
is going to trust a luminosity estimate which is based solely on beam
diagnostics; the luminosity spectrum must come from interaction data.
Nothing else can reflect all of the fluctuations which might take
place in the detailed dynamics of beam-beam collision as well as the
bunch to bunch variations of linac behaviour.

The job can not be done by counting small angle ($30<\theta <120 mr$)
Bhabha scattering events, as we do at LEP or SLC.  The only way of
knowing the energy of such events would be by direct calorimetric
measurement of the electrons -- inside the mask.  But even the best
electromagnetic calorimetry could do little better than 1\%, which is
not good enough, and even that will be hard to achieve with the severe
background and shower-containment problems in the mask region.

\begin{table}
\begin{center}
\begin{tabular}{|l|c|l|} \hline
 Process & Rate & Comment \\
 \hline \hline
 Bhabha 180--300 mr  & 223R & Endcap. Best statistics, adequate
 procistion \\
\hline
 Bhabha 300--800 mr  & 104R & Intermediate \\
\hline
 Bhabha 800--2341 mr  & 8R & Barrel. Lower statistics, good precision \\
\hline
 $\mu^+\mu^-$ & R & Low statistics, good precistion \\ 
\hline
 $Z^0\gamma$ & 30R & Reasonable statistics. Should study further \\
\hline
 $W^+W^-$  & 12R &  Reasonable statistics.  Poor precision\\
\hline
 Two real $\gamma$ & 2R & Low statistics, reasonable precision \\
\hline \hline
 $t-\bar t$ & $\sim$R & Signal\\ 
\hline
\end{tabular}
\caption{Approximate rates for possible luminosity and measurement processes
        }
\label{table:processes}
\end{center}
\end{table}

Table~\ref{table:processes} shows some of the possible processes which
might be used for luminosity monitoring.  The ideal process would have
a rate which is much higher than any of the interesting physics
processes - ``many units of R'' - and the final state tracks should be
sufficiently well measured in the main part of the detector so that
the event energy can be obtained to better than 1/1000 by fitting the
angles.  Nothing fully satisfies these criteria.  Radiative return to
the $Z^0$ ($e^+ e^- \rightarrow Z^0 \gamma$) is one of the interesting
channels because the final state gamma ray will usually be unseen and
can be constrained to lie within some tens of milliradians from the
beam direction. The angles of the $Z^0$ decay products then give a
good measure of the energy.  But only 7\% of the 30 units of R will
have $Z^0 \rightarrow e^+ e^-$ or $\rightarrow \mu^+ \mu^-$ which can
be well measured.  It will be much harder to get accurate directions
for the jets from the hadronic $Z^0$ decays, especially since they will be
strongly boosted towards the endcap regions where detectors will be
less uniform and some jet energy will be lost into the masks.
Nevertheless, this will be an important channel, especially because
the mass of the $Z^0$ is so well measured, giving an absolute energy
scale quite independently from measurements on the beams.  Good
calculations of the radiative corrections to $Z^0$ ($e^+ e^-
\rightarrow Z^0 \gamma $) must be on the shopping list \shop.

\begin{figure}[htb]
 \begin{center}
   \epsfig{file=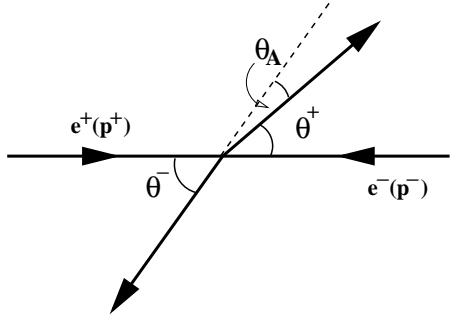,width=0.5\linewidth} 
 
  \vspace{-0.2cm}
  \caption{\label{fig:acoll} 
        Definition of the acollinearity angle $\theta_A$.}
 \end{center}
\end{figure}

My recommended best solution to the problem is to use large-angle 
Bhabha scattering events, with electrons going into the endcap regions 
of the detector.  The rates are large because the t-channel exchange 
still dominates, and there are only two clean lepton tracks to 
measure, with well determined angles in a properly designed detector .
There is, of course, no way of fitting these events to get individual 
event energies $\sqrt{s}$ from angles alone.  But we can measure a 
different quantity for each event, the acollinearity angle between the
two electrons, $\theta_A$, see Figure~\ref{fig:acoll}.  The
luminosity spectrum can be unfolded from the distribution of
$\theta_A$.  This was first demonstrated analytically in
1992~\cite{frarymill} and has since been checked
numerically~\cite{kurihara}, though a full Monte Carlo simulation has
not yet been done.

If the spread of the beam energies at collision were simply Gaussian
then it is easy to show that
\begin{displaymath}
\sigma_{\sqrt{s}} = \sigma_{\Delta p} = (\sigma_{\theta_A}
p_b)/\sin{\theta},
\end{displaymath}
where, for a small momentum mismatch $\Delta p$ between the two
interacting particles, $\Delta p = p^+ - p^-$, $\sqrt{s} = p^+ + p^-$,
$p^+ \simeq p^- \simeq p_b$ and $\theta^+ \simeq \theta^- \simeq
\theta$.  The $\sin{\theta}$ on the bottom line means that the method
becomes more precise at larger angles, but there is a trade-off with
the rate which means that the endcap region $150<\theta <500 mr$ is
likely to be the most useful, see Table~\ref{table:processes}.

The actual distribution of beamstrahlung losses is very nongaussian,
and it combines with the inescapable initial state radiation (ISR)
losses for individual events to give spectra for the colliding
electrons which have a strong peak at close to the nominal beam energy
and a long tail due to radiation.  From the {\em signed} acollinearity
distribution we can separately extract the shape of both the $e^+$ and
$e^-$ momentum distributions at collision and hence calculate the
distribution of $\sqrt{s}$.  It is important to reconstruct correctly
the spike of events with very small radiation losses because these
give the sharpness to the threshold step in the excitation curve from
which the mass can be measured.  So we must resolve the acollinearity
to better than 1/1000; not an impossible job for a good tracking
detector, so long as the endcap regions are designed with it in mind.
In the 1992 study~\cite{frarymill} we found that approximately 20\% of
events were in the spike with $\theta_A < 1 mr$ for the best linac
designs, after beamstrahlung, ISR and linac beamspread had been
considered.

Cross sections for Bhabha scattering at more than 150 mr have never
been calculated as precisely as those at small angles, where they are
needed for LEP luminosity measurements and where systematic errors are
now dropping below 1 per mille.  If the high-luminosity TESLA gets
built then we will have physics samples of $500fb^{-1}$ with $~10^6$
events in some channels, so it is time to put this region of Bhabha
scattering onto the shopping list for precision calculations.  The
cross sections will be needed to normalise high statistics channels
\shop, and good acollinearity distributions from initial state
radiation will be required for the luminosity spectrum measurement
\shop.

\section{Light Higgs}
This summer's fits~\cite{ewfit} to electroweak data tell us that the
if the Higgs boson has properties close to the standard model it
should either be seen at LEP ($m_h<105$ GeV) or early in the LHC
programme.  But discovering it will just be the beginning of Higgs
physics.  It will then be necessary to establish what kind of Higgs
boson we have seen; standard model, minimal supersymmetric extension,
next to minimal extension, etc. etc.  The linear collider will be
where these questions are answered.  Measuring the Yukawa coupling to
top quarks, Section~\ref{sec-top} above, is one of the tests, but the
most important goal is to determine the total width of the Higgs and
its partial widths for different quark, gluon and boson flavours. A
worthwhile step towards this will be the measurement of the total rate
for $e^+ e^- \rightarrow Z^0 h$, where the $Z^0$ decays to $e^+ e^-$
or $\mu^+ \mu^-$ and the missing mass of the recoiling Higgs boson is
measured without relying on any of the Higgs decay
products~\cite{cdrphys}.  The resolution required on the two outgoing
lepton tracks in this process dictates the precision needed in the
tracking detectors at the Future Linear Collider.  Within this Higgs
sample it is then necessary to identify particular sets of final
states.  With $50fb^{-1}$ at $\sqrt{s}=350$ GeV and $m_h = 140$ GeV the
1996 ECFA/DESY study showed that in its standard detector the product
$\sigma_{Zh}\cdot BR(h \rightarrow b \bar{b})$ could be measured to 6.2\%
but $\sigma_{Zh}\cdot BR(h \rightarrow c \bar{c} + gg)$ would have an
estimated error of 47\%, clearly not good enough.  The extra
statistics at the high luminosity version of TESLA would be very
important for this measurement.  The Snowmass
detector~\cite{snowmass}, with its smaller beampipe radius and good
microvertex system, aims to separate the $c \bar{c}$ from the $gg$
events.  All of these processes must be on the phenomenlogists'
shopping lists for more refined calculations \shop.

\section{The Compton Collider}
\begin{figure}[htb]
 \begin{center}
   \epsfig{file=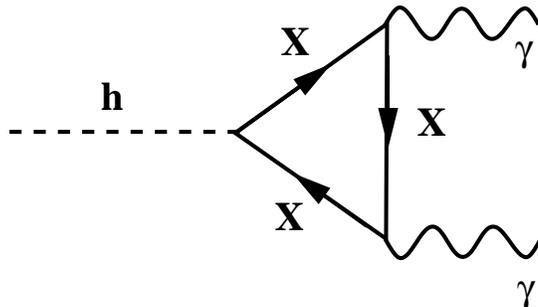,width=0.6\linewidth} 
 
  \vspace{-0.4cm}
  \caption{\label{fig:loop} 
        Higgs boson coupling to $\gamma\gamma$. All loop particles of
  the same kind (heavy quarks, heavy leptons, bosons, squarks, etc.)
  contribute equally so long as $m_X>100$ GeV.}
 \end{center}
\end{figure}

To measure the absolute width of the Higgs boson may require the
Compton Collider~\cite{gamgam} mode of the linac.  From Higgs decays
at the LHC or at the Linear Collider we should be able to measure the
branching ratio to $\gamma \gamma$.  Then with a known real two-photon
flux we can measure the partial width $\Gamma^h_{\gamma \gamma}$ for
$m_h<350$ GeV.  This quantity is also of great interest in its own
right because it gets finite and comparable loop contributions, see
Figure~\ref{fig:loop}, from every heavy charged particle which couples
to the Higgs boson, with no upper mass limit~\cite{djmorioka}.  It is
very important to check all of the calculations of these cross
sections and branching ratios \shop.  If the Linear Collider is built
after a light Higgs has already been discovered there may be an
overwhelming case to install a second interaction region with a
backscattered laser photon facility from the start. 

If the $\gamma \gamma$ luminosity can be made comparable with the $e^+
e^-$ luminosity of the linac then the production rate of $W^+ W^-$
will actually be higher in the $\gamma \gamma$ mode, and a 
complementary set of processes can be studied.  Even if the luminosity 
is substantially lower than for $e^+ e^-$, scattering electrons from 
real photons in the $e \gamma$ mode will give by far the best way of 
measuring the structure of the photon~\cite{djmvogt}.   But already at
LEP we are limited in the precision of our analysis by inadequacies in
the available Monte Carlo models that have to be used in unfolding the
structure functions~\cite{djmpic}.  More work is needed from the
phenomenologists, especially on the right way to treat the $c \bar{c}$
threshold \shop.

\section{SUSY}
\begin{figure} 
\vspace*{-1.2cm}
\begin{center}
\hspace*{-11mm}
\epsfig{file= 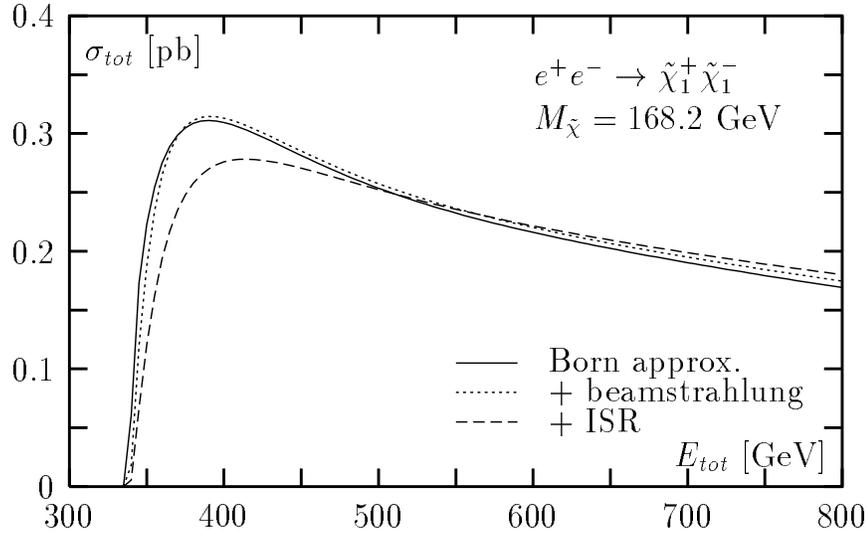,width=13.cm} \\
\vspace*{-0.3cm}
\hspace*{-7mm}
\epsfig{file= 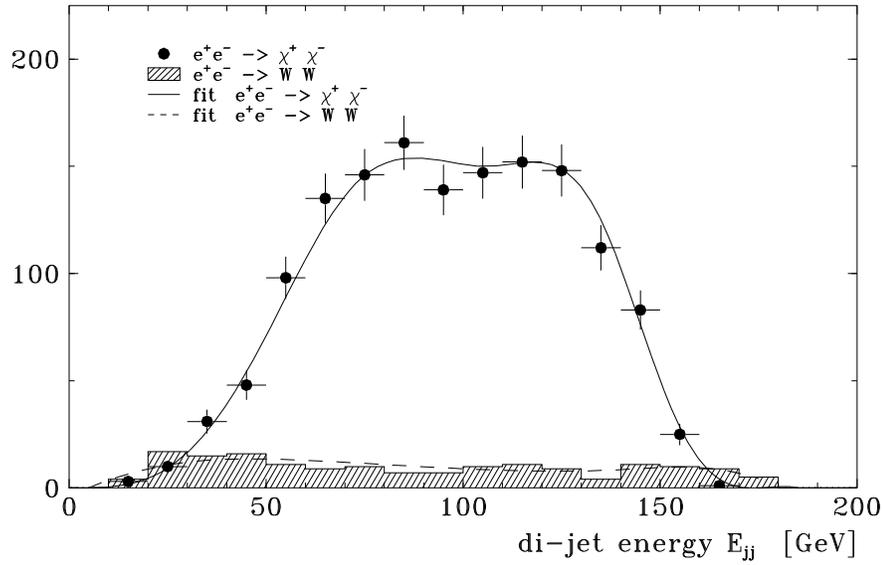,width=8.5cm,angle=90}
\end{center}
\vspace{-0.5cm}
\caption[]{
  Upper part: Threshold behavior of the cross section for $e^+e^-
  \rightarrow \tilde{\chi}^+_1 \tilde{\chi}^-_1$ including
  initial-state radiation and beamstrahlung.
  Lower part: Simulation of the energy spectrum in the decay
  $\tilde{\chi}^+_1 \rightarrow \tilde{\chi}^0_1 + jj$ based on the
  input values $m_{\tilde{\chi}^+_1} = 168.2$~GeV and
  $m_{\tilde{\chi}^0_1} = 88.1$~GeV at $\sqrt{s}$ = 500 GeV;
  Ref. 10.  \label{fig:chi+}}
\end{figure}

There is an infinite set of shopping lists \shop\ in SUSY because there 
are so many possible combinations of -- as yet unconstrained -- 
parameters, but the circumstantial evidence for the existence of 
Supersymmetry continues to grow.  If SUSY has been discovered at the 
LHC before the Linear Collider is built there will still be the 
question ``which SUSY model is this?''.  The LHC may already have seen 
some coloured SUSY particles, and part of the Higgs sector, but the 
Linear Collider would be needed to find the charginos and sleptons.  And
the well defined kinematics of $e^+ e^-$ collisions will allow 
threshold scans, or precise mass determinations from the spectra of
recoil particles.  The 1996 ECFA/DESY Study~\cite{cdrphys} found, for
instance, that in $50fb^{-1}$ at $\sqrt{s}= 500$ GeV the mass of a $170$
GeV $\tilde{\chi}^+$ could be measured to $\pm 100$ MeV, see
Figure~\ref{fig:chi+}.  We are promised polarised electrons and
perhaps polarised positrons at the same time.  How can these be used
to probe the subtleties of the theory? \shop

At recent workshop meetings the consequences of the alternative
gauge-mediated SUSY-breaking scheme have been
discussed~\cite{ambrosanio}, including the possibility that long-lived
sparticles will decay some tens of centimetres into the the detectors.
To make good measurements on events like this would require the
capacity to point-back gamma rays from the calorimeter with very good
angular precision.  Should we be trying to do this in the first stage
of Linear Collider operation? \shop

\section{QCD} 
A sample of $500fb^{-1}$ at $\sqrt{s}= 500$ GeV, for a high luminosity
collider, will have $~10R \simeq
4pb$ of $q\bar{q}$, giving 2 million events with two and more final
state jets, comparable to the samples studied at LEP on the $Z^0$ peak
where the errors on $\alpha_s$ were systematics limited. With such
statistics it would be possible to make a significant step in charting
the running of $\alpha_s$ up to a much higher scale. Phenomenology
must follow \shop.  There might also be an opportunity to use the
radiative return sample $e^{+}e^{-} \rightarrow Z^0 \gamma$ with $Z^0 
\rightarrow q\bar{q} \rightarrow $jets to repeat measurements on the
$Z^0$ in the same detector, cancelling out some of the systematics
involved in a comparison with LEP data.  But these $Z^0$s would be
highly boosted into the endcap regions of the detector.  If the
physics case for such measurements is strong enough there are two
alternative strategies for producing an unboosted $Z^0$ sample in the
same detector; either the machine can be adapted for running at the
$Z^0$ energy, or it might be run with asymmetric energies to give a
sample of radiative-return events with reduced forward boost.  Again,
better models \shop\ of the radiative return will be needed.

\section{Electroweak Physics}
\begin{figure}[htb]
 \begin{center}
   \epsfig{file=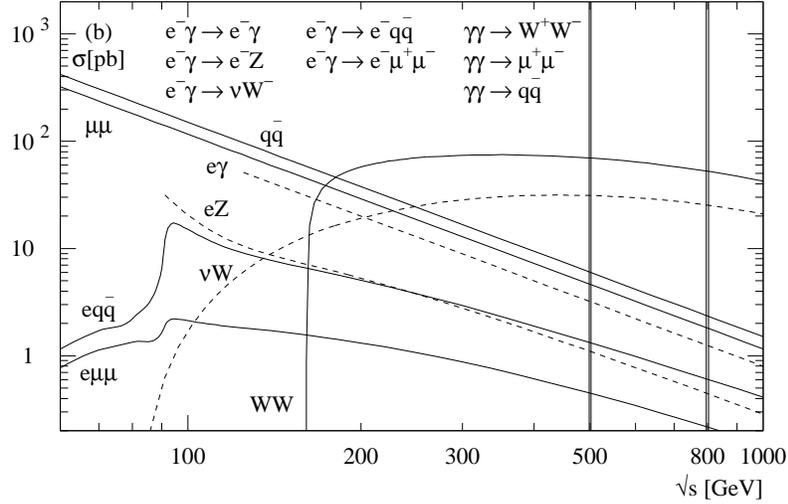,width=0.9\linewidth,clip=} 
 
  \vspace{-0.4cm}
  \caption{\label{fig:gg} 
        Cross section for some $e\gamma$ and $\gamma\gamma$
  processes. Note that some of them (e.g. to $WW$) are bigger than
  the equivalent $e^+e^-$ cross sections (Fig 1).}
 \end{center}
\end{figure}

Study of the $W^+ W^-$ final state is a growth industry at LEP2.
There will be even more sensitivity to anomalous couplings at the
linear collider: the increased energy will change the balance between
the three tree level graphs, giving better intrinsic sensitivity; the
greater luminosity (especially at TESLA) will more than compensate for
the falling cross section; and it should be possible to polarise both
beams to alter the balance of the graphs in a controlled way \shop.
Also, at $\sqrt{s}= 500$ GeV, the single W rate
grows to the same size as $W^+ W^-$, giving
a new handle on the couplings.  
Figure~\ref{fig:gg} shows cross sections for processes which can be
studied in the $e\gamma$ and $\gamma\gamma$ modes of the machine, some
of which are  sensitive~\cite{boudappi,djmorioka} 
to different electroweak couplings \shop.
Some of these cross sections
grow with $\sqrt{s}$ so they will be even more important at 1 TeV.

\section{Conclusions} 
\begin{enumerate}
\item A linear collider, especially with high luminosity, will do much
      new precision physics which will be complementary to the LHC.
\item Many of the processes involved need better calculations than are
      at present available, and better Monte Carlo generators.
\item Theoretical input is needed soon for decisions which have to be
      taken about: 
 \begin{itemize} 
    \item Precision of detectors, 
    \item Need for polarised $e^-$ {\em and $e^+$}, 
    \item Importance of the $e^-e^-$, $\gamma e$ 
          and $\gamma \gamma$ options, 
    \item Need to measure $h \rightarrow c \bar{c}$, 
          requiring small beampipe, low background etc.  
    \item How to measure $\Gamma^{total}_{higgs}$.
 \end{itemize}
\item Phenomenologists are encouraged to join their local linear
      collider workshops (see webpages~\cite{ECFAD,asia,america,wcoord}). 
\end{enumerate}

\section{Acknowledgements}
I am grateful to collegues in the second ECFA/DESY study~\cite{ECFAD} 
for access to their results. Dr. Jan Lauber's help with the figures is
greatly appreciated.

\clearpage

\noindent{\bf References}


\begin{thebibliography}{99}
%
\bibitem{ECFAD} Webpages for 2nd ECFA/DESY Study on Physics and
Detectors at a Future Linear Collider at
``http://www.desy.de/conferences/ecfa-desy-lc98.html''
\bibitem{asia} Webpages for the ACFA Asian activities at
``http://acfahep.kek.jp''
\bibitem{america} Webpages for the American activities at\\
``http://lcwws.physics.yale.edu/lc/america.html''
\bibitem{wcoord} Webpages for the Worldwide co-ordination committee at\\
``http://lcwws.physics.yale.edu/lc/''
\bibitem{morioka} Proceedings of the Workshop on Physics and
Experiments with Linear Colliders, Morioka-Appi, Japan, September
1995, eds A.~Miyamoto, Y.~Fujii, T.~Matsui, S.~Iwata, World
Scientific, Singapore (1996)
\bibitem{waikoloa} Proceedings of the Workshop on Physics and
Experiments with Linear Colliders, Wiakoloa, Hawaii, April 1993, eds
F.A.~Harris, S.L.~Olsen, S.~Pakvasa, X.~Tata, World Scientific,
Singapore (1993)
\bibitem{saariselka} Proceedings of the Workshop on Physics and
Experiments with Linear Colliders, Saariselka, Finland, September
1991, eds R.~Orava, P.~Eerola, M.~Nordberg, World Scientific,
Singapore (1992)
\bibitem{desy123} DESY Laboratory reports ``123 series'':--\\ 
Vols A and B, $e^+ e^-$
Collisions at 500 GeV, The Physics Potential, (DESY 92-123A, DESY
92-123B) edited by P.M.~Zerwas (1992)\\ Vol C, $e^+ e^-$ Collisions at
500 GeV, The Physics Potential, (DESY 93-123C) edited by P.M.~Zerwas
(1993)\\ Vol D, $e^+ e^-$ Collisions at TeV Energies, The Physics
Potential, (DESY 96-123D) edited by P.M.~Zerwas (1996)\\ Vol E, $e^+
e^-$ Linear Colliders, Physics and Detector Studies, (DESY 97-123E)
edited by R.~Settles (1997)
\bibitem{techrev}International Linear Collider Technical Review
Report, edited by G.Loew, SLAC-R-95-471 (1995)
\bibitem{tesla} Conceptual Design of a 500 GeV $e^+ e^-$ Linear
Collider with Integrated X-ray Laser Facility, 2 volumes, editors
R.~Brinkmann, G.~Materlik, J.~Rossbach, A.~Wagner, DESY 1997-048 (or
ECFA 1997-182)
\bibitem{gamgam}I.~Ginzburg, G.~Kotkin, V.~Serbo,V.~Telnov, Pizma
ZhETF, {\bf 34}, 514 (1981)\\ I.~Ginzburg, G.~Kotkin,
V.~Serbo,V.~Telnov, NIM, {\bf 205}, 47 (1983)\\
V.~Telnov, Principles of Photon Colliders, in Proceedings of the
Workshop on Gamma-Gamma colliders, LBL, Berkeley, March 28-31 1994;
eds S.Chattopadhyay and A.M.Sessler, NIM {\bf
A355}, 1-194 (1995)
\bibitem{djmorioka}D.J.Miller, Other Options: $e^- e^-$, $e \gamma$
and $ \gamma \gamma$ Physics at a Linear Collider, ibid~\cite{morioka}
pp305-321
\bibitem{fermiweb} Information on physics at a muon collider is at the
website: http://fnphyx-www.fnal.gov/conferences/femcpw97/workshop.html
\bibitem{Keilrol}J.~Ellis, E.~Keil, L.~Rolandi; Options for Future
Colliders at CERN . CERN-EP-98-003 ; CERN-TH-98-033 .
(CERN-SL-98-004-AP)
\bibitem{Blondel}A.~Blondel, these proceedings.
\bibitem{Xbandweb}Webpages for Stanford NLC programme;\\
``http://www-project.slac.stanford.edu/nlc/index.html''
\bibitem{JLC}Webpages for KEK JLC programme; ``http://lcdev.kek.jp/''
\bibitem{CLIC}Webpages for the CERN CLIC project;\\
``http://www.cern.ch/CERN/Divisions/PS/CLIC/Welcome.html''
\bibitem{snowmass}Webpages for the Snowmass study;\\
``http://fnpx03.fnal.gov/conferences/snowmass96/''
\bibitem{jlc}JLC-I, KEK Report 92-16 (1992)
\bibitem{hoang} A.H.~Hoang, T.~Teubner; Phys.Rev. D58 (1998) 114023
\bibitem{melnikov} K.~Melnikov, A.~Yelkhovsky; Nucl.Phys. B528 (1998)
59-72
\bibitem{Fujii}K.~Fujii; quoted by P.~Igo-Kemenes,
ibid~\cite{waikoloa} 117.
\bibitem{juste}A.~Juste; contributions to ECFA/DESY workshops, see e.g. \\
``http://www.lal.in2p3.fr/Workshop/ECFA-DESY-LC98/slides/webrama/aaaaavtc/dia1.htm''
\bibitem{dittm} S.~Dittmaier, M.~Kr\"amer, Y.~Liao, M.~Spira,
P.M.~Zerwas; Higgs Radiation off Top Quarks in $e^+e^-$ Collisions:
hep-ph/9808433
\bibitem{frarymill}M.~Frary, D.J.~Miller; Monitoring the Luminosity
Spectrum, DESY 92-123A 379-391
\bibitem{kurihara}Y.~Kurihara, contribution on behalf of the KEK JLC
group to First ECFA/DESY workshop, Munich, September 1996.
\bibitem{ewfit}See LEP Electroweak Working Group webpages;\\
``http://www.cern.ch/LEPEWWG/''
\bibitem{cdrphys}E.~Accomando {\em et al}, (Chapter 1 of
ibid~\cite{tesla}) Physics Reports 299(1) (1998) 1-78
\bibitem{djmvogt}D.J.~Miller, A.~Vogt; Kinematic Coverage for
determining the Photon Structure Function $F^{\gamma}_2$,
DESY 96-123D 473-478.
\bibitem{djmpic} D.J.~Miller; Photon Structure and gamma-gamma
Physics; In proceedings of International Physics in Collision, June
1998, ed F.Fabbri; hep-ex/9807017
\bibitem{ambrosanio}S.~Ambrosanio; talk at Lund Workshop of 2nd
ECFA/DESY Study, 28 June 1998, see webpages\\
``http://www.quark.lu.se/workshop/ECFA/slides/default.htm''
\bibitem{boudappi}F.~Boudjema; Aspects of W Physics at the Linear
Collider, ibid~\cite{morioka} 199-227.
\end{thebibliography}
\end{document}